\documentclass[12pt]{article}
\usepackage{psfrag,amsmath,epsfig,a4,cite,latexsym,axodraw}

\def\pslash#1{{\setbox0=\hbox{$#1$}
  \rlap{\ifdim\wd0>.7em\kern.22\wd0\else\kern.1\wd0\fi /}#1}}

\def\ghat{{\hat{g}}}
\def\gtilde{{\tilde{g}}}

\def\ttbar{t\bar{t}}
\def\DRED{{\scshape dred}}
\def\DREG{{\scshape dreg}}
\def\mDRED{{\rm \scriptscriptstyle DRED}}
\def\mDREG{{\rm \scriptscriptstyle DREG}}
\def\Amp#1#2#3{{\cal A}_{{\rm #3}}^{#2}{({\scriptstyle{#1}})}}
\def\ASq#1#2#3{{\cal M}_{{\rm #3}}^{#2}{({\scriptstyle{#1}})}}
\def\MAv#1#2#3{\langle{\cal M}_{\rm #3}^{#2}{({\scriptstyle{#1}})}\rangle}
\def\ngRS{n_G^{\rm \scriptscriptstyle RS}}
\def\ng{n_g^{\mDRED}}
\def\nphi{n_\phi^{\mDRED}}

\allowdisplaybreaks[1]
\begin{document}
\begin{flushright}
IPPP/05/52\\
DCPT/05/104\\
{\tt hep-ph/0508203}\\
\end{flushright}
\vspace{3em}
\begin{center}
{\Large\bf Factorization and Regularization by\\[1ex]
 Dimensional Reduction}
\\
\vspace{3em}
{\sc Adrian Signer, Dominik St\"ockinger%
\footnote{email: \{Adrian.Signer, Dominik.Stockinger\}@durham.ac.uk}%
}\\[2em]
{\it Institute for Particle Physics Phenomenology,\\ University of Durham,
Durham DH1~3LE, UK}
\end{center}
\vspace{2ex}
\begin{abstract}
Since an old observation by Beenakker et al, the evaluation of QCD
processes in dimensional reduction has repeatedly led to terms that
seem to violate the QCD factorization theorem. We reconsider the
example of the process $gg\to\ttbar$ and show that the
factorization problem can be completely resolved. A natural
interpretation of the seemingly non-factorizing terms is found, and
they are rewritten in a systematic and factorized form. The key to the
solution is that the $D$- and $(4-D)$-dimensional parts of the
4-dimensional gluon have to be regarded as independent partons.
\end{abstract}


\section{Introduction}

Nearly 20 years ago, Ref.\ \cite{BKNS} observed a problem concerning
factorization in conjunction with regularization by dimensional
reduction (\DRED) \cite{Siegel79}. The partonic process $gg\to
\ttbar$ with non-vanishing quark mass $m_t\equiv m$ was evaluated
using both \DRED\ and ordinary dimensional regularization (\DREG)
\cite{HV}%
\footnote{Sometimes, these regularization schemes are also abbreviated
as ``DR'' and ``CDR (conventional dimensional regularization)''.
}%
. 
Contrary to expectations \cite{Factorization}, the difference between
the two regularization schemes could not be absorbed by a finite
additional factorization, corresponding to a change in the parton
distribution functions. 

As a consequence it seems impossible to write the hadronic
cross section ${\sigma}_{\rm had}$ as a convolution of parton
distribution functions $f$ and the partonic cross 
section $\sigma_{\rm parton}^{\mDRED}$. Schematically, the
factorization problem can be expressed as
\begin{equation}
\sigma_{\rm had} = f \otimes \sigma_{\rm parton}^{\mDRED}
+\mbox{extra, non-factorizing terms}.
\end{equation}
The extra terms vanish in the limit of vanishing quark mass $m=0$.
Moreover, in the case of only massless partons the transition between
the two regularization schemes has been worked out for many examples
\cite{KST,CST} and could always be performed as expected. Even for
one-loop amplitudes involving massive partons the transition could be
studied \cite{CDT}. Nevertheless, the problem found in Ref.\
\cite{BKNS} for the massive case has remained unsolved. It has
repeatedly shown up and has been stressed again e.g.\ in Refs.\
\cite{BHZ96,vNS}.

There are two areas where \DRED\ is traditionally applied with great
benefit. One is the evaluation of purely massless amplitudes,
especially within QCD. Here \DRED\ and related methods allow the use of
powerful helicity methods \cite{ManganoParke}. The most important
application of \DRED\ and its original purpose is the regularization 
of supersymmetric theories. It has been shown to preserve
supersymmetry relations in many different cases at the one-loop
\cite{CJN80,BHZ96,STIChecks} and the two-loop level
\cite{DREDPaper,Harlander05}, and in \cite{DREDPaper} also further
properties such as mathematical consistency have been established.

The factorization problem is particularly troublesome for the
calculation of QCD corrections to supersymmetric processes involving
hadrons. In spite of the advantages of \DRED, it renders the use of
\DRED\ questionable (see e.g.\ the discussion in
\cite{BHZ96}). Resorting to \DREG\ in such calculations introduces 
several disadvantages. Mainly, supersymmetry is broken and
has to be restored by adding supersymmetry-restoring counterterms
\cite{BHZ96,STIChecks} that do not correspond to multiplicative
renormalization. In addition, the $\overline{DR}$ renormalization
scheme, a very common definition of supersymmetry parameters, is
naturally based on \DRED\ but only awkward to realize using \DREG.
Clearly, a resolution of the factorization problem would be welcome
for both fundamental and practical reasons \cite{SPA,Kalinowski}.

In this article we reconsider the problem found in Refs.\
\cite{BKNS,vNS}. We show that, despite first appearances, the result
of Ref.\ \cite{BKNS} in fact is perfectly consistent with
factorization. 

We begin in Sec.\ \ref{sec:calculation} with a detailed explanation of
the calculation of the LO process $gg\to \ttbar$ and the real
NLO correction $gg\to \ttbar g$. We consider the collinear limit of
two of the gluons and recover the seemingly paradoxical result of
Ref.\ \cite{BKNS}. An important ingredient of this collinear limit is
the necessity to average over the unobservable azimuthal angle of the
final state gluon. It distinguishes the massive from the massless
case, and in the massive case it leads to the difference between the
\DREG- and the \DRED-result.

In Sec.\ \ref{sec:solution} we first describe the general idea that
will lead to a re-interpretation of the result, showing that it is
consistent with factorization. The crucial point to notice is that in
\DRED\ the 4-dimensional gluon is a composition of a $D$-dimensional
part and a remaining $(4-D)$-dimensional part, and that these two
parts behave as {\em two different partons} $g$ and $\phi$.

Finally it is demonstrated in detail how this idea leads to a
resolution of the factorization problem. On the one hand, in the
collinear limit the NLO cross section becomes equal to a linear
combination of {\em two   different} LO cross sections, with either
$g$ or $\phi$ in the initial state. On the other hand, the appearing
prefactors in this linear combination have a natural interpretation as
splitting functions for the splitting processes $g\to gg$, $\phi\to
g\phi$, etc. We will also explain why factorization works in the $m=0$
case already without distinguishing between $g$ and $\phi$.

In Sec.\ \ref{sec:conclusions} we give our conclusions.

\section{Recovering the seemingly non-factorizing result}
\label{sec:calculation}

\subsection{LO and NLO calculation}

We consider hadroproduction of a quark pair $\ttbar$ via gluon
fusion, the process for which the factorization problem has been
reported in Refs.\ \cite{BKNS,vNS}. In 
this section we will briefly describe the required tree-level
calculations and recover the result of these references. At leading
order (LO) we only need the $2\to 2$ process $gg\to \ttbar$, whereas
at next-to-leading order (NLO) we also have to consider the
$2\to3$ process with an additional gluon in the final state.

We carry out the calculation using either \DREG\ or \DRED. In both cases,
space-time, momenta and momentum integrals are treated in $D$
dimensions. In \DREG, the gluon vector field is treated in $D$
dimensions as well, while in \DRED\ the gluon field and
$\gamma$-matrices remain 4-dimensional quantities.

\begin{figure}
\centerline{
\begin{picture}(130,90)(0,0)
\SetWidth{.6}
\put(0,75){$k_1$}
\put(0,0){$k_2$}
\put(100,75){$p_1$}
\put(100,0){$p_2$}
\Gluon(15,70)(45,50){3}{4} 
\Gluon(15,10)(45,30){3}{4}
\ArrowLine(65,50)(95,70)
\ArrowLine(95,10)(65,30)
\GCirc(55,40){15}{.85}
\end{picture}
\hfill
\begin{picture}(130,90)(0,0)
\SetWidth{.6}
\put(0,75){$k_1$}
\put(0,0){$k_2$}
\put(100,75){$p_1$}
\put(110,40){$p_2$}
\put(100,0){$k_3$}
\Gluon(15,70)(45,50){3}{4} 
\Gluon(15,10)(45,30){3}{4}
\ArrowLine(65,50)(95,70)
\ArrowLine(103,40)(70,40)
\Gluon(95,10)(65,30){3}{4}
\GCirc(55,40){15}{.85}
\end{picture}
\hfill
\begin{picture}(130,90)(0,0)
\SetWidth{.6}
\put(0,75){$k_1$}
\put(0,15){$k_2$}
\put(80,2){$k_3$}
\put(100,75){$p_1$}
\put(100,20){$p_2$}
\Gluon(14,70)(54,52){3}{5.5} 
\Gluon(34,33)(54,42){3}{2}
\Gluon(14,24)(34,33){3}{2}
\Gluon(76,5)(34,33){3}{6}
\Vertex(34,33){2}
\ArrowLine(72,55)(96,71)
\ArrowLine(96,23)(72,39)
\GCirc(64,47){12}{.85}
\end{picture}
}
\caption{
\label{fig:diagrams}
Generic structure of (a) LO diagrams, (b) NLO real correction
diagrams, (c) NLO diagrams giving rise to a collinear divergence for
$k_3\to(1-x)k_2$.}
\end{figure}
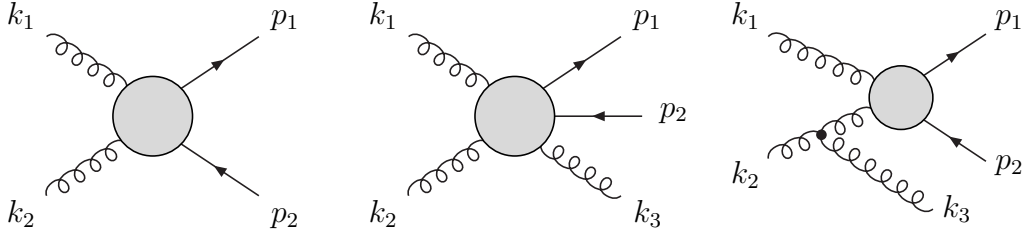

At leading order the amplitude $\Amp{2\to 2}{}{RS}$ is
given by the diagrams sketched in Fig.\ \ref{fig:diagrams}a. The
subscript ${\rm RS}$ denotes the regularization scheme, \DREG\ or
\DRED. The incoming gluon momenta and colour indices are denoted by
$k_{1,2}$ and $a_{1,2}$, respectively; the outgoing momenta are called
$p_{1,2}$. We will use the kinematical variables
\begin{align}
S&=2k_1 k_2,&
T_1&=(k_1-p_1)^2-m^2,&
U_1&=(k_2-p_1)^2-m^2.
\end{align}
$\Amp{2\to 2}{}{RS}$ can be decomposed into two colour
structures as
\begin{equation}
\Amp{2\to 2}{}{RS} = \Amp{2\to 2}{(12)}{RS}
 T^{a_1}T^{a_2} + \Amp{2\to 2}{(21)}{RS} T^{a_2}T^{a_1}.
\end{equation}
The squared LO amplitude, summed over initial and final state
polarizations and colours, can be decomposed as
\begin{eqnarray}
\lefteqn{\ASq{2\to 2}{}{RS} = 
    \sum_{\rm pols, col}|\Amp{2\to 2}{}{RS}|^2 } &&
\nonumber \\
&& =\frac{(N^2-1)^2}{4N} \ASq{2\to 2}{(1)}{RS}
-\frac{N^2-1}{4N} \ASq{2\to 2}{(2)}{RS},
\end{eqnarray}
where $N=3$ is the number of colours. For the polarization sum
corresponding to a gluon with polarization vector $\epsilon^\mu$ and
momentum $k$, we use
\begin{equation}
\sum_{\rm pols}\epsilon^\mu\epsilon^\nu{}^* \to
-g^{\mu\nu}+\frac{n^\mu k^\nu+k^\mu n^\nu}{(nk)}
-\frac{n^2 k^\mu k^\nu}{(nk)^2}
\label{polsum}
\end{equation}
with an arbitrary gauge vector $n^\mu$ such that $nk\ne0$.

We obtain the following results:
\begin{subequations}
\begin{align}
\ASq{2\to 2}{(1,2)}{RS}  &= 8g^4\left\{1-\frac{2T_1U_1}{S^2},\
                           +\frac{2T_1U_1}{S^2}\right\}B_{\rm QED},\\
B_{\rm QED} &=\ngRS\left(-1+\frac{\ngRS S^2}{4T_1U_1}\right)
+\frac{4m^2 S}{T_1^2U_1^2}\left(T_1U_1-m^2 S\right),
\label{LOresult}
\end{align}
\end{subequations}
in agreement with Ref.\ \cite{BKNS}. The difference between the
calculation in \DRED\ and \DREG\ enters only through the number
$\ngRS$ of gluon degrees of freedom,
\begin{equation}
n_G^{\mDREG}=D-2,\quad
n_G^{\mDRED}=2.
\end{equation}
Technically, $\ngRS$ appears in the form $\ngRS=g^\mu_\mu-2$, where
the metric tensor originates either from the numerator of a gluon
propagator or the polarization sum (\ref{polsum}).

At NLO, we restrict ourselves to the real corrections, corresponding
to the process $gg\to\ttbar g$. This is sufficient for the discussion
of the collinear divergences and the factorization problem
\cite{BKNS,vNS}. The diagrams contributing to the amplitude
$\Amp{2\to 3}{}{}$ are generically depicted in Fig.\
\ref{fig:diagrams}b. The outgoing momentum and colour indices of the
additional final state gluon are denoted by $k_3$, $a_3$; in
accordance with Ref.\ \cite{BKNS} we use the kinematical variables
\begin{equation}
\begin{aligned}
s&=(k_1+k_2)^2,&
s_4&=(k_3+p_1)^2-m^2,&
t'&=(k_2-k_3)^2,\\
u'&=(k_1-k_3)^2,&
u_6&=(k_2-p_1)^2-m^2,&
u_7&=(k_1-p_1)^2-m^2,
\end{aligned}
\end{equation}
which satisfy $s+s_4+t'+u'+u_6+u_7=0$. It is useful to keep the
distinction between these variables for the $2\to3$ process and the
variables $S$, $T_1$, $U_1$ for the $2\to2$ process although $S$ and
$s$ are the same functions of $k_{1,2}$. The explicit form of the full
result for $\ASq{2\to 3}{}{RS}$ is lengthy and suppressed here.

In order to obtain the partonic cross sections, the squared amplitudes
have to be averaged over the initial state polarizations
and colours and divided by a flux factor. We denote these averaged
quantities by
\begin{subequations}
\label{MAvDef}
\begin{align}
\MAv{2\to2}{}{RS} &= \frac{1}{2S}\frac{1}{[\ngRS (N^2-1)]^2}
\ASq{2\to 2}{}{RS},\\
\MAv{2\to 3}{}{RS} &= \frac{1}{2s}\frac{1}{[\ngRS (N^2-1)]^2}
\ASq{2\to 3}{}{RS}.
\end{align}
\end{subequations}
The differential cross sections are then given by ($P\equiv
k_1+k_2-p_1-p_2$)
\begin{align}
d\sigma_{2\to 2}^{\rm RS} & = \MAv{2\to 2}{}{RS}
\,\left(\prod_{p_f=p_{1,2}}\frac{d^{D-1}p_f}{2p_f^0(2\pi)^3}\right)
\,(2\pi)^D\delta^{(D)}\left(P\right)
,
\\
d\sigma_{2\to 3}^{\rm RS} & = \MAv{2\to 3}{}{RS}
\,\left(\prod_{p_f=p_{1,2},k_3}\frac{d^{D-1}p_f}{2p_f^0(2\pi)^3}\right)
\,(2\pi)^D\delta^{(D)}\left(P-k_3\right)
.
\end{align}
They depend on the regularization scheme at ${\cal O}(4-D)$ and at
${\cal O}((4-D)^0)$ due to soft and collinear divergences.

\subsection{Collinear limit and azimuthal average}

Now we consider the limit of $\MAv{2\to 3}{}{RS}$, where the
unobserved final state gluon becomes collinear to one of the initial
state gluons. To be specific we will concentrate on the collinear
limit $2\|3$ of gluon 2 and gluon 3 and define the collinear limit
$k_\perp\to 0$ by parametrizing the momenta $k_2^\mu$ and $k_3^\mu$ as
follows:
\begin{equation}
\begin{aligned}
\label{kcoll}
k_3^\mu &= (1-x) k_2^\mu+k_\perp^\mu - 
\frac{k_\perp^2}{1-x} \frac{n^\mu}{2 k_2 n},
\end{aligned}
\end{equation}
where the auxiliary vector $n^\mu$ satisfies
$n^2 = n k_\perp=0$.

The collinear divergence in the NLO cross section originates from
diagrams of the form shown in Fig.\ \ref{fig:diagrams}c where the
virtual gluon becomes on-shell. In the squared amplitude this gives
rise to terms of the order $1/t'\sim 1/k_\perp^2$, and such terms lead
to singularities in the phase-space integral. As can be read off from
Fig.\ \ref{fig:diagrams}c, one would expect the divergent NLO terms to
become proportional to the LO terms with the identification
\begin{align}
\label{coll23LO}
S&\to xs,&
U_1&\to x u_6,&
T_1&\to -x(s+u_6).
\end{align}
However, this naive expectation does not take into account the
following subtlety: not all poles $1/k_\perp^2$ of the squared
amplitude are directly of the form $1/t'$. Some poles have a more
involved structure. In particular, in our example, there are poles of
the form $(s s_4-u' u_6)^2/t'{}^2$.  Upon taking the collinear limit,
these terms depend on $k_\perp^\mu$. However, the transverse direction
$k_\perp^\mu$ is unobservable in the collinear limit and will be
azimuthally averaged over in the phase-space integral~\cite{CST}.
This averaging procedure affects only terms containing $1/t'{}^2$, and
it yields
\begin{align}
\frac{(ss_4-u'u_6)^2}{t'{}^2}
\ & \stackrel{\langle 2\|3\rangle}{\longrightarrow}\ 
\frac{1}{D-2} \frac{-(1-x)}{x^2} 
\frac{4 S  \left(T_1 U_1  - m^2 S \right)}{t'} + \ldots\ ,
\end{align}
where the dots denote terms without a $1/t'$ singularity. The notation
$\langle2\|3\rangle$ implies that the average over the
$(D-2)$-dimensional transverse space is taken in the collinear limit.
The factor $(D-2)$ enters the denominator as a result of this
averaging~\cite{CST}.

Taking the averaged collinear limit of $\MAv{2\to 3}{}{RS}$ we
obtain 
\begin{align}
\MAv{2\to 3}{}{RS}  
\stackrel{\langle2\|3\rangle}{\longrightarrow} 
\frac{-4 g^2 N}{t'} \,
&\Bigg(
\frac{(1-x+x^2)^2}{x(1-x)} \,
\MAv{2\to 2}{}{RS} 
\nonumber\\
&+\Big(\frac{\ngRS}{D-2}-1\Big)
\frac{(1-x)}{x}\MAv{2\to 2}{}{RS}|_{m}
\Bigg)
,
\label{CollDRED}
\end{align}
where $\ASq{2\to 2}{}{RS}|_m \equiv \ASq{2\to 2}{}{RS}- \ASq{2\to
2}{}{RS}|_{m=0}$ denotes the mass terms of $\ASq{2\to 2}{}{RS}$.  This
equation is equivalent to the result found in Refs.\ \cite{BKNS,vNS}.

The factorization theorem seems to suggest that the terms that are
divergent in this collinear limit are proportional to the LO
result. Whereas the first term on the right-hand side of
(\ref{CollDRED}) is in accordance with this expectation, the second term
contains only the mass-dependent terms of the LO result and,
therefore, seems to violate the factorization theorem. Due to the
prefactor, this second term is absent in \DREG, and the problem is only
present in \DRED. What we would expect in going from \DREG, where
factorization holds, to \DRED\ is a change in the function multiplying
the LO term, but not a change in the structure of the result.

As mentioned in Ref.~\cite{BKNS,vNS} the problematic term 
vanishes in the massless limit. However, this is not generally true
but is peculiar to the process under consideration. The decisive
feature is not the mass of the quarks but the presence of terms $\sim 
1/t'{}^2$. In our case, the absence of $1/t'{}^2$ terms in the
massless case can be explained by helicity conservation.

In the next section we will discuss the origin of the seemingly
non-factorizing term and show that it can be rewritten in a way that
is consistent with factorization.

\section{Reconciling the NLO result with factorization}
\label{sec:solution}

\subsection{General idea}
\label{sec:idea}

In the collinear limit, the NLO result in \DRED\
(\ref{CollDRED}) does not seem to factorize into a product of a
splitting function and the LO result. In contrast, the NLO result in
\DREG\ does factorize. There is a simple argument that allows to
understand why the two regularization schemes behave in such a
different way.
In the regularized
expressions, the number of dimensions $D$ and of gluon degrees of
freedom $\ngRS$ can be set to integers. 
For example, \DREG\ with integer $D$ and $n_G^{\mDREG}=D-2$
simply corresponds to unregularized QCD in $D$ dimensions. Of course,
factorization can be expected to hold in QCD with an arbitrary number
of dimensions. This is the reason why Eq.\ (\ref{CollDRED}) factorizes
in the case of \DREG.


In contrast, \DRED\ with e.g.\ $D=3$ does not lead to 3-dimensional QCD
but rather to 4-dimensional QCD, dimensionally reduced to 3
dimensions. It is well known that in the process of dimensional
reduction from 4 to 3 dimensions, the 4-dimensional gluon is
decomposed into the 3-dimensional gluon $A^\mu$ $(\mu=0,1,2)$ and
an extra scalar field $\phi\equiv A^3$. The resulting theory is
3-dimensional QCD, supplemented with a minimally coupled scalar $\phi$
in the adjoint representation.

The crucial point to be learnt from this discussion is that the
dimensionally reduced theory contains {\em two distinct partons}, the 
3-dimensional gluon $g$ and the scalar $\phi$. At LO there are
therefore four distinct partonic processes for $\ttbar$ production:
\begin{align}
gg&\to\ttbar,&
g\phi&\to\ttbar,&
\phi g&\to\ttbar,&
\phi\phi&\to\ttbar.
\label{fourLOprocesses}
\end{align}
It is obvious that factorization can be expected to hold in this
dimensionally reduced theory, but not in the same way as in \DREG.
On the right-hand side of Eq.\ (\ref{CollDRED}) we do not expect one
single term but instead a linear combination of all four partonic LO
processes.

In \DRED\ with arbitrary, non-integer $D$, the situation is
similar. The regularized theory contains a $D$-dimensional gluon $g$
and $4-D$ additional scalar fields $\phi$, so-called
$\epsilon$-scalars \cite{CJN80}. Again, $g$ and $\phi$ have to be
viewed as independent partons, and the collinear limit is expected to
contain all four LO processes.

In order to express this new expectation more formally, we denote the
4-dimensional gluon that has always been assumed in \DRED\ in the
previous section (and also in Refs.\ \cite{BKNS,vNS}) by $G$. The
$2\to2$ and $2\to3$ processes considered in the previous section can
then be written more explicitly as
\begin{align}
\ASq{2\to2(3)}{}{\mDRED}&\equiv\ASq{GG\to\ttbar(G)}{}{\mDRED}.
\end{align}
Since the 4-dimensional gluon $G$ constitutes the combination
$g+\phi$, the squared matrix elements satisfy the relation
\begin{align}
\ASq{GG\to\ttbar}{}{\mDRED}=&
\ASq{gg\to\ttbar}{}{\mDRED}+
\ASq{g\phi\to\ttbar}{}{\mDRED}
\nonumber\\+&
\ASq{\phi g\to\ttbar}{}{\mDRED}+
\ASq{\phi\phi\to\ttbar}{}{\mDRED}
\label{Ggphirelation}
\end{align}
and similarly for $\ASq{GG\to\ttbar G}{}{\mDRED}$. This leads us to
expect that the collinear limit in \DRED\ can be written as
\begin{align}
\MAv{ij\to\ttbar k}{}{\mDRED} 
\stackrel{\langle2\|3\rangle}{\longrightarrow}&
\frac{-2g^2}{t'}\Bigg[
\sum_{l=g,\phi}P_{j\to lk}
\MAv{il\to\ttbar}{}{\mDRED}
\Bigg].
\label{newDREDexpectation}
\end{align}
Contrary to the corresponding formula for \DREG,  the
right-hand side of  Eq.~(\ref{newDREDexpectation}) is a
linear combination involving more than one LO process.

In the following we will show that the seemingly non-factorizing term
in Eq.~(\ref{CollDRED}) can be rewritten as a linear combination of
the four partonic LO processes. Thus, factorization is valid in \DRED\
in the form expected in Eq.~(\ref{newDREDexpectation}) and we will
see that the functions $P_{j\to l k}$ can be interpreted as splitting
functions.
 
\subsection{Collinear limit and LO result with $g$ or $\phi$ in the
  initial state}

According to the idea discussed in the preceding subsection we
evaluate all four partonic LO processes (\ref{fourLOprocesses})
individually.  The algebraic 
expressions for the partonic processes involving $g$, $\phi$, or $G$
are distinguished by the values of the polarization vector
$\epsilon^\mu$ and the corresponding polarization sum. The
polarization sum corresponding to an external $G$ is the one given in
Eq.\ (\ref{polsum}); the ones corresponding to $g$ and $\phi$ read
\begin{subequations}
\label{polsumsgphi}
\begin{align}
g:\quad&
\sum_{\rm pols}\epsilon^\mu\epsilon^\nu{}^* \to
-\ghat^{\mu\nu}+\frac{n^\mu k^\nu+k^\mu n^\nu}{(nk)}
-\frac{n^2 k^\mu k^\nu}{(nk)^2},
\\
\phi:\quad&
\sum_{\rm pols}\epsilon^\mu\epsilon^\nu{}^* \to
-\gtilde^{\mu\nu}.
\end{align}
\end{subequations}
The objects $\ghat^{\mu\nu}$ and $\gtilde^{\mu\nu}$ are the projectors
on the $D$- and $(4-D)$-dimensional subspaces \cite{Siegel79} (see
also Ref.\ \cite{DREDPaper} for further details) and satisfy
$\ghat^{\mu\nu}\ghat_{\mu\nu} =D$, $\gtilde^{\mu\nu}\gtilde_{\mu\nu}
=4-D$ and the projector relations $g^{\mu\nu}\ghat_\nu{}^\rho
=\ghat^{\mu\rho}$, $g^{\mu\nu}\gtilde_\nu{}^\rho
=\gtilde^{\mu\rho}$. They are related to the 4-dimensional metric
tensor by $g^{\mu\nu} = \ghat^{\mu\nu}+\gtilde^{\mu\nu}$.

We obtain the following results:
\begin{subequations}
\label{LOgphi}
\begin{align}
\ASq{ij\to\ttbar}{(1,2)}{\mDRED}  &= 8g^4\left\{1-\frac{2T_1U_1}{S^2},\
                           +\frac{2T_1U_1}{S^2}\right\}B_{ij},\\
\label{ggresult}
B_{gg} &=\ng\left(-1+\frac{\ng S^2}{4T_1U_1}\right)
+\frac{4m^2 S}{T_1^2U_1^2}\left(T_1U_1-m^2 S\right),\\
B_{\phi\phi} &=\nphi\left(-1+\frac{\nphi S^2}{4T_1U_1}\right),\\
B_{g\phi} &=\ng\nphi\frac{ S^2}{4T_1U_1},\\
B_{\phi g}&=\ng\nphi\frac{ S^2}{4T_1U_1}.
\end{align}
\end{subequations}
Here the symbols $\ng$ and $\nphi$ denote the numbers of degrees of
freedom corresponding to the partons $g$ and $\phi$:
\begin{equation}
\ng=D-2,\quad
\nphi=4-D.
\end{equation}
The fact that the 4-dimensional gluon $G$ is the combination of $g$
and $\phi$ is reflected in the equality $n_G^{\mDRED}=\ng+\nphi$ and
by the observation that, as already stated in Eq.\
(\ref{Ggphirelation}), the sum of the four partial results
(\ref{LOgphi}) is equal to the result for the $GG$ initial state.

Note that the result for the $gg$ case is equal to the LO result in
\DREG\ because eqs.\ (\ref{LOresult}) and (\ref{ggresult}) have the
same form and $n_G^{\mDREG}=\ng$. This equality can be understood as a
consequence of the fact that in the simple process $gg\to\ttbar$ at
tree level no $\epsilon$-scalars $\phi$ appear as virtual states in
the Feynman diagrams.

In a next step we perform the calculation of all eight squared
amplitudes $\ASq{ij\to \ttbar k}{}{\mDRED}$ with $i,j,k=g,\phi$.  We
do not present the full analytic results but concentrate on the
collinear limit $k_3\to(1-x)k_2$, since we are interested in how the
processes involving the individual partons $g$, $\phi$ behave as
compared to the seemingly non-factorizing result (\ref{CollDRED}) for
the process involving only $G$. The averaged amplitudes are defined as
in Eq.\ (\ref{MAvDef}), replacing $n_G^{\mDRED}$ by $\ng$, $\nphi$
where appropriate. We find the following results: 
\begin{subequations}
\label{Collgphi}
\begin{align}
\MAv{ig\to\ttbar g}{}{\mDRED}&\stackrel{\langle2\|3\rangle}{\longrightarrow}
\frac{-4g^2 N}{t'}\MAv{ig\to\ttbar}{}{\mDRED}\,\frac{(1-x+x^2)^2}{x(1-x)},\\
\MAv{i\phi\to\ttbar g}{}{\mDRED}&\stackrel{\langle2\|3\rangle}{\longrightarrow}
\frac{-4g^2 N}{t'}\MAv{i\phi\to\ttbar}{}{\mDRED}\,\frac{x}{1-x},\\
\MAv{ig\to\ttbar \phi}{}{\mDRED}&\stackrel{\langle2\|3\rangle}{\longrightarrow}
\frac{-4g^2 N}{t'}\MAv{i\phi\to\ttbar}{}{\mDRED}\,\frac{\nphi }{\ng}\,x(1-x),\\
\MAv{i\phi\to\ttbar \phi}{}{\mDRED}
&\stackrel{\langle2\|3\rangle}{\longrightarrow}
\frac{-4g^2 N}{t'}\MAv{ig\to\ttbar}{}{\mDRED}\,\frac{1-x}{x}.
\end{align}
\end{subequations}
These results have precisely the form of Eq.\
(\ref{newDREDexpectation}) with
\begin{subequations}
\label{Pdred}
\begin{align}
P_{g\to gg} &=
2N\,\frac{(1-x-x^2)^2}{x(1-x)},\\
P_{\phi\to \phi g} &=
2N\,\frac{x}{1-x},\\
P_{g\to \phi \phi} &=
2N\,\frac{\nphi }{\ng}\,x(1-x),\\
P_{\phi\to g\phi} &=
2N\,\frac{1-x}{x},\\
P_{j\to lk} &= 0 \mbox{ otherwise}.
\end{align}
\end{subequations}
They demonstrate clearly that all eight individual partonic
processes factorize in the usual way into a product of a splitting
function and a LO process, without any unusual terms. There are not
even non-trivial linear combinations of LO processes on the right-hand
sides. This fact and the origin of the splitting functions is
discussed in the following subsections.

The eight results can now be combined to reconcile the collinear limit
in Eq.~(\ref{CollDRED}) for \DRED\ with factorization. Instead of Eq.\
(\ref{CollDRED}) we now obtain
\begin{align}
\MAv{GG\to\ttbar G}{}{\mDRED}
\hspace{-0em}&\hspace{0em}
=\sum_{i,j,k=g,\phi}
\frac{n_i^{\mDRED}n_j^{\mDRED}}{(n_G^{\mDRED})^2}
\MAv{ij\to\ttbar k}{}{\mDRED}
\nonumber\\[1ex]
\stackrel{\langle2\|3\rangle}{\longrightarrow}
\ 
\frac{-4g^2 N}{t'}
\Bigg[
&\MAv{Gg\to\ttbar}{}{\mDRED}
\bigg(\frac{\ng}{n_G^{\mDRED}}\,
\frac{(1-x+x^2)^2}{x(1-x)}+
\frac{\nphi}{n_G^{\mDRED}}
\,\frac{1-x}{x}\bigg)
\nonumber\\
+&\MAv{G\phi\to\ttbar}{}{\mDRED}
\,\frac{\nphi}{n_G^{\mDRED}}\,\bigg(
\frac{x}{1-x}
+x(1-x)\bigg)
\Bigg],
\label{CollDRED2}
\end{align}
where the relations $\ASq{Gj\to\ttbar}{}{}=\ASq{gj\to\ttbar}{}{}
+\ASq{\phi j\to\ttbar}{}{}$ have been used. In this equation the
collinear limit finally acquires a factorized structure although
\DRED\ is used. As expected in Sec.\ \ref{sec:idea}, a linear
combination of LO processes appears on the right-hand side.

It is instructive to directly verify the equality of
eqs.~(\ref{CollDRED2}) and (\ref{CollDRED}), the factorized and
non-factorized version of the collinear limit respectively.  Since
the mass dependence in Eq.~(\ref{LOgphi}) enters only through the $gg$
result we can write
\begin{subequations}
\label{MassTerms}
\begin{align}
\label{MassTermRes}
\MAv{GG\to\ttbar}{}{\mDRED}|_m &= \frac{\ng}{n_G^{\mDRED}}\left(
\MAv{Gg\to\ttbar}{}{\mDRED}-\MAv{G\phi\to\ttbar}{}{\mDRED}\right),\\
\label{MasslessRes}
\MAv{GG\to\ttbar}{}{\mDRED}|_{m=0}&=\MAv{Gg\to\ttbar}{}{\mDRED}|_{m=0}=
\MAv{G\phi\to\ttbar}{}{\mDRED}.
\end{align}
\end{subequations}
Thus we see that the disturbing mass term in Eq.~(\ref{CollDRED})
indeed can be resolved as a linear combination of complete LO
processes. Using eqs.\ (\ref{MassTerms}) in Eq.\ (\ref{CollDRED})
directly leads to Eq.\ (\ref{CollDRED2}). 

Finally we note that in the massless case (\ref{MasslessRes}), several
of the LO processes become equal, which is why the collinear limit
then takes a simpler form and the problematic term in Eq.\
(\ref{CollDRED}) disappears. This is however a peculiarity of the
considered process and related to the absence of terms $\sim 1/t'{}^2$
discussed in Sec.\ \ref{sec:calculation}, but it is not a generic
feature of processes with massless partons.

\subsection{Splitting functions involving $g$ and $\phi$}

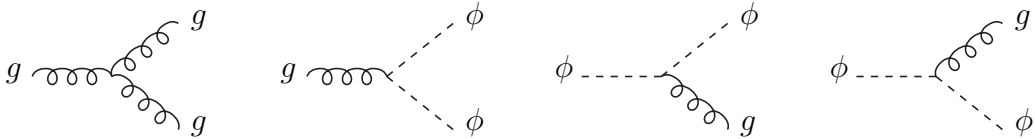
\begin{figure}[t]
\begin{picture}(100,60)(0,0)
\SetWidth{.6}
\put(10,30){$g$}
\put(80,50){$g$}
\put(80,10){$g$}
\Gluon(20,30)(50,30){3}{3} 
\Gluon(50,30)(75,50){3}{3}
\Gluon(50,30)(75,10){3}{3}
\end{picture}
\begin{picture}(100,60)(0,0)
\SetWidth{.6}
\put(10,30){$g$}
\put(80,50){$\phi$}
\put(80,10){$\phi$}
\Gluon(20,30)(50,30){3}{3} 
\DashLine(50,30)(75,50){3}
\DashLine(50,30)(75,10){3}
\end{picture}
\begin{picture}(100,60)(0,0)
\SetWidth{.6}
\put(10,30){$\phi$}
\put(80,50){$\phi$}
\put(80,10){$g$}
\DashLine(20,30)(50,30){3}
\DashLine(50,30)(75,50){3}
\Gluon(50,30)(75,10){3}{3}
\end{picture}
\begin{picture}(100,60)(0,0)
\SetWidth{.6}
\put(10,30){$\phi$}
\put(80,50){$g$}
\put(80,10){$\phi$}
\DashLine(20,30)(50,30){3}
\Gluon(50,30)(75,50){3}{3}
\DashLine(50,30)(75,10){3}
\end{picture}
\caption{
\label{fig:splitting}
Tree level diagrams for the four splitting processes involving $g$ and
$\phi$.
}
\end{figure}

In this subsection we focus on the splitting functions appearing in
Eq.~(\ref{Pdred}), involving $g$ and $\phi$ as partons. In order to
consolidate our understanding of factorization in \DRED\ we will
present an independent derivation of these splitting
functions. Instead of reading them off from the collinear limits of
particular NLO processes we directly evaluate the amplitudes for the
splitting processes
\begin{equation}
\begin{aligned}
g&\to g(x)\,g(1-x),&
g&\to \phi(x)\,\phi(1-x),\\
\phi&\to \phi(x)\, g(1-x),&
\phi&\to g(x)\,\phi(1-x).
\end{aligned}
\end{equation}
The corresponding diagrams are shown in Fig.~\ref{fig:splitting}. Note
that the amplitudes for splitting processes involving an odd number of
$\phi$ partons vanish at tree level. In each splitting process $i\to
jk$ the momenta are assigned as $p_i\equiv k_2$, $p_k\equiv k_3$ as
given in Eq.\ (\ref{kcoll}), and $p_j=k_2-k_3$. In order to obtain the
splitting probabilities, the amplitudes are squared and summed over
colours and polarizations according to Eq.\ (\ref{polsumsgphi}). Only
particle $j$ is kept slightly off-shell, $p_j^2\sim k_\perp^2$, and its
Lorentz and colour indices are kept uncontracted. The result for each
splitting process thus has the form ${\cal P}^{\rho\rho',aa'}_{i\to
jk}$, where $\rho$, $\rho'$ and $a$, $a'$ are the open Lorentz and
colour indices.  Terms subleading in $k_\perp$ are neglected
and the average over the $D-2$ transverse directions is performed.
Finally, terms proportional to $(k_2-k_3)^\rho$ or $(k_2-k_3)^{\rho'}$
can be neglected, too. Due to the Ward identity they do not contribute
if the splitting processes are part of a larger physical process where
all particles are on-shell.

After these manipulations, the results ${\cal P}^{\rho\rho',aa'}_{i\to
jk}$ take the form
\begin{align}
\label{CollOp}
{\cal P}^{\rho\rho',aa'}_{i\to jk} = 
P_{i\to jk}(x)\; \frac{2(k_2-k_3)^2}{x}\;
\frac{n^{\mDRED}_i}{n^{\mDRED}_j }\;
\delta_{aa'} \; g_{(j)}^{\rho\rho'}
.
\end{align}
They are proportional to
$\ghat^{\rho\rho'}$ if $j=g$ and to $\gtilde^{\rho\rho'}$ if $j=\phi$
(commonly abbreviated as $g_{(j)}^{\rho\rho'}$ here), and they are
proportional to $\delta_{aa'}$ in colour space. As expected, the
prefactors are given by the splitting functions $P_{i\to jk}(x)$ of 
Eq.~(\ref{Pdred}), multiplied by additional factors that compensate
for the different prefactors in cross sections with either $i$ or $j$
in the initial state.

Hence the functions given in Eq.~(\ref{Pdred}) have a natural
interpretation as universal splitting functions. The fact that only
one term appears on each right-hand side of Eq.\ (\ref{Collgphi}) is
due to the vanishing of the splitting functions involving an odd
number of $\phi$'s.

For future reference we introduce splitting functions corresponding
to 4-dimensional gluons
\begin{align}
n_G^{\mDRED} P_{G\to j G} &=
\sum_{k=g,\phi}\left(\ng P_{g\to j k}+\nphi P_{\phi\to jk}
\right)
\label{PGjG}
\end{align}
and note that the splitting functions satisfy the sum rule
\begin{align}
n_G^{\mDRED} P_{g\to gg} &= \sum_{j=g,\phi}n_G^{\mDRED} P_{G\to j G}
=\sum_{i,j,k=g,\phi}n_i^{\mDRED} P_{i\to j k}.
\label{sumrule}
\end{align}
As in Eq.~(\ref{CollOp}), the factors $n_G^{\mDRED}$, $\ng$ and
$\nphi$ appear because we are considering the splitting of an initial
state parton and, therefore, have to correct for the factors due to
the average over polarizations.

We close the subsection with several remarks. First, note that
$P_{g\to gg}$ is identical to the well-known gluon splitting 
function in \DREG. Second, the splitting functions involving $\phi$
coincide with the splitting functions involving massless squarks and
gluons, given in Ref.\ \cite{CDT}, if the colour factors for squarks
$T_R$, $C_F$ are replaced by $C_A=N$. The particular splitting
function $P_{g\to\phi\phi}$ has already been made use of in Ref.\
\cite{CST} in order to study the difference between \DRED\ and
\DREG. And finally, $P_{\phi\to g\phi}$ is the prefactor of 
the puzzling term in Eq.\ (\ref{CollDRED}), and it corresponds to the
factor $K_g$ in Ref.\ \cite{vNS}. The nature of $P_{\phi\to g\phi}$ as
a splitting function explains the universal behaviour of $K_g$
described in this reference.

\subsection{Final result}

In the previous subsections we have seen that the real NLO processes
with partons $g$, $\phi$ indeed factorize in the collinear limit. The
$x$-dependent prefactors can be interpreted as the splitting functions
$P_{i\to jk}$ corresponding to the parton splittings $g\to gg$, $g\to
\phi\phi$, $\phi\to g\phi$, $\phi\to\phi g$. Thus the results for the
collinear limits take a very systematic form: 
\begin{align}
\MAv{ij\to\ttbar k}{}{\mDRED} 
\stackrel{\langle2\|3\rangle}{\longrightarrow}&
\frac{-2g^2}{t'}\Bigg[
\sum_{l=g,\phi}P_{j\to lk}
\MAv{il\to\ttbar}{}{\mDRED}
\Bigg],
\label{Collgphi2}
\end{align}
where $i,j,k=g,\phi$. The sums on the right-hand side all collapse to
one single term since only the four aforementioned splitting functions
can contribute, while splitting functions with an odd number of
$\phi$'s vanish at tree level. Similarly, using the combinations
(\ref{PGjG}) of splitting functions involving $G$, the result for the
process involving only 4-dimensional gluons can be expressed as
\begin{align}
\MAv{GG\to\ttbar G}{}{\mDRED} \
\stackrel{\langle2\|3\rangle}{\longrightarrow}\ 
\frac{-2g^2}{t'}
\Bigg[&\sum_{j=g,\phi}
P_{G\to j G}
\MAv{Gj\to\ttbar}{}{\mDRED}
\Bigg].
\label{CollDRED3}
\end{align}
Although there is a non-trivial structure on the right-hand side, this
result has the form that is expected from the factorization theorem. 

For comparison, we repeat the corresponding result for the case of
\DREG\ with adapted notation:
\begin{align}
\MAv{gg\to\ttbar g}{}{\mDREG} \
\stackrel{\langle2\|3\rangle}{\longrightarrow}\ 
\frac{-2g^2}{t'}
\Bigg[&
P_{g\to g g}
\MAv{gg\to\ttbar}{}{\mDREG}
\Bigg].
\end{align}

In the remainder of this section we briefly discuss the relation
between \DRED\ and \DREG\ for the cross sections. The results for the
collinear limits can be elevated to the level of cross sections by
performing the suitable phase-space integration and taking into
account the second collinear limit $1\|3$. The singular terms in the
collinear limits yield the subtraction terms that render the cross
section finite. In \DREG, the subtracted hard scattering cross section
$d\hat{\sigma}^{\mDREG}$ at NLO is given by 
\begin{align}
\int d\hat{\sigma}^{\mDREG}_{gg\to\ttbar g} = 
\int & d\sigma^{\mDREG}_{gg\to\ttbar g}
\nonumber\\
+
\Bigg[
&\int_0^{1-\delta} dx_1
\left(\frac{\alpha_s}{2\pi}\frac{1}{\epsilon}P_{g\to gg}(x_1)\right)
d\sigma^{\mDREG}_{gg\to\ttbar}(x_1k_1,k_2)
\nonumber\\
+
&\int_0^{1-\delta} dx_2
\left(\frac{\alpha_s}{2\pi}\frac{1}{\epsilon}P_{g\to gg}(x_2)\right)
d\sigma^{\mDREG}_{gg\to\ttbar}(k_1,x_2k_2)\Bigg]
\label{DREGSigmaHat}
\end{align}
with $\alpha_s=g^2/(4\pi)$ and $D=4-2\epsilon$.
In \DRED\ it can be defined analogously:
\begin{align}
\int d\hat{\sigma}^{\mDRED}_{GG\to\ttbar G} = 
\int & d\sigma^{\mDRED}_{GG\to\ttbar G}
\nonumber\\
+
\sum_{j=g,\phi}
\Bigg[
&\int_0^{1-\delta} dx_1
\left(\frac{\alpha_s}{2\pi}\frac{1}{\epsilon}P_{G\to jG}(x_1)\right)
d\sigma^{\mDRED}_{jG\to\ttbar}(x_1k_1,k_2)
\nonumber\\
+
&\int_0^{1-\delta} dx_2
\left(\frac{\alpha_s}{2\pi}\frac{1}{\epsilon}P_{G\to jG}(x_2)\right)
d\sigma^{\mDRED}_{Gj\to\ttbar}(k_1,x_2k_2)
\Bigg].
\label{DREDSigmaHat}
\end{align}
In these equations, all integration regions are assumed to contain the
same collinear regions. The small parameter $\delta>0$ excludes the
region around $x_i=1$, which would lead to further infrared
singularities that cancel only by adding the virtual NLO corrections.

These subtracted cross sections are free of collinear singularities
and, by construction, the non-singular remainders in both
regularization schemes are equal up to terms of ${\cal O}(4-D)$:%
\footnote{Note that the factorization scheme has been implicitly
  fixed in eqs.\ (\ref{DREGSigmaHat}), (\ref{DREDSigmaHat}). Different
  factorization schemes can be realized by adding identical terms in
  the brackets multiplying the $d\sigma_{ij\to\ttbar}$ in eqs.\
  (\ref{DREGSigmaHat}), (\ref{DREDSigmaHat}). The resulting subtracted
  cross sections in \DREG\ and \DRED\ are then still equal.}
\begin{equation}
\int d\hat{\sigma}^{\mDREG}_{gg\to\ttbar g} = 
\int d\hat{\sigma}^{\mDRED}_{GG\to\ttbar G}+{\cal
  O}(4-D)
.
\label{SigmaHatEq}
\end{equation}
Eqs.\ (\ref{DREGSigmaHat})--(\ref{SigmaHatEq}) can also be derived
directly from the puzzling result Eq.\ (6.28) in Ref.\ \cite{BKNS} by
inserting our expression (\ref{MassTermRes}) for the disturbing mass
term. 

This shows that the final hadronic cross section, which is obtained
from $d\hat\sigma$ through convolution with parton distribution
functions, can be evaluated both using \DREG\ or using \DRED. In
particular, eq.\ (\ref{SigmaHatEq}) shows that the same factorization
scheme can be realized using either \DREG\ or \DRED, and therefore the
same parton distribution functions (e.g.\ defined in the
$\overline{MS}$ factorization scheme) have to be used in both cases.
The structure of the calculation is the same. The only difference is
the appearance of the two independent partons $g$, $\phi$ in the
subtraction terms for \DRED\ that lead from $d\sigma$ to
$d\hat\sigma$.

\section{Conclusions}
\label{sec:conclusions}

We have considered the factorization problem of \DRED\ that has
repeatedly shown up in the literature \cite{BKNS,BHZ96,vNS}. Eq.\
(\ref{CollDRED}) exhibits the seemingly non-factorizing terms in the
collinear limit of the process $gg\to\ttbar g$. We have shown that the
problem can be completely solved. 

The key to the solution is to consider the 4-dimensional gluon $G$ in
\DRED\ as a combination of the $D$-dimensional gluon $g$ and $4-D$
$\epsilon$-scalars $\phi$. If $g$ and $\phi$ are treated as
independent partons as  in Eq.\ (\ref{CollDRED2}), the collinear limit
acquires a factorized form. The problematic terms on the right-hand
side are replaced by a linear combination of several LO processes
involving $g$ and $\phi$. Furthermore we have shown that the
coefficients in this linear combination have a natural interpretation
as splitting functions.

The final form of the collinear limit is displayed in eqs.\
(\ref{Collgphi2}) and (\ref{CollDRED3}). We have shown that the result
for the collinear limit can be transferred to the level of cross
sections and that the hadron cross section can be evaluated using both
\DREG\ or \DRED. All results have a very systematic and natural
structure. 

In summary, the factorization problem of \DRED, i.e.\ the presence of
seemingly non-factorizing terms, is not a problem but a signal that
the distinction between $g$ and $\phi$ as independent partons cannot
be ignored. The solution does not affect the computation of the NLO
diagrams itself. Only the expectation from the collinear limit and the
structure of the subtraction terms needed to obtain the hard
scattering cross section have to reflect this distinction. Although we
have only considered the process $gg\to\ttbar g$ as an example and
ignored virtual NLO corrections, one can expect that factorization in
\DRED\ holds in general and even in higher orders.  The details of the
general construction of finite, regularization-independent hard
scattering cross sections will be left for future work.

\smallskip\smallskip

An interesting remaining question is for which processes the
factorization problem and the decomposition of the 4-dimensional gluon
as $G=g+\phi$ is relevant in general. While a general answer to this
question is beyond the scope of the present article, we can give two
criteria, based on the analysis of the considered process, where the
problem disappears for $m=0$.

From the point of view of Sec.\ \ref{sec:calculation}, for $m=0$ the
terms of the order $1/t'{}^2$ vanish. In this case, no average over
the transverse direction of the collinear gluon has to be performed.
Therefore, the result in \DRED\ is trivially the $D=4$ limit of the
\DREG-result, and in both regularizations factorization holds in the
naive way.

From the point of view of Sec.\ \ref{sec:solution}, in the massless
case the LO processes with $GG$, $Gg$ or $G\phi$ in the initial state
all become equal, see Eq.\ (\ref{MasslessRes}). As a result, in the
collinear limit (\ref{CollDRED3}) no distinction between the different
LO processes has to be made, and the prefactors combine to the sum
$P_{G\to gG}+P_{G\to \phi G}$, which is simply equal to $P_{g\to gg}$
according to the sum rule (\ref{sumrule}). Hence the collinear limit
in \DRED\ again reduces to the naive form involving only 4-dimensional
gluons and one splitting function $P_{g\to gg}$.

The situation is different for the process with one more leg,
$gg\to\ttbar g$ with a hard gluon in the final state. We have checked
that for this example, e.g.\ $\MAv{Gg\to\ttbar G}{}{\mDRED}
\ne\MAv{G\phi\to\ttbar G}{}{\mDRED}$ already for $m=0$ in contrast to
Eq.\ (\ref{MasslessRes}). Therefore, the factorization problem is not
generally linked to the presence of massive partons but rather to 
sufficiently complicated kinematics. 

\subsection*{Acknowledgements}

We are grateful to P.\ Marquard, T.\ Plehn, W.\ Porod,
A.\ Vogt, and P.\ Zerwas for 
useful discussions.

\begin{flushleft}

\end{flushleft}

\end{document}